# The use of Virtual Reality in Enhancing Interdisciplinary Research and Education

Tiffany LEUNG
School of Computing, Queen's University
Kingston, ON, Canada

Farhana ZULKERNINE
School of Computing, Queen's University
Kingston, ON, Canada

Haruna ISAH
School of Computing, Queen's University
Kingston, ON, Canada

## ABSTRACT

Virtual Reality (VR) is increasingly being recognized for its educational potential and as an effective way to convey new knowledge to people, it supports interactive and collaborative activities. Affordable VR powered by mobile technologies is opening a new world of opportunities that can transform the ways in which we learn and engage with others. This paper reports our study regarding the application of VR in stimulating interdisciplinary communication. It investigates the promises of VR in interdisciplinary education and research. The main contributions of this study are (i) literature review of theories of learning underlying the justification of the use of VR systems in education, (ii) taxonomy of the various types and implementations of VR systems and their application in supporting education and research (iii) evaluation of educational applications of VR from a broad range of disciplines, (iv) investigation of how the learning process and learning outcomes are affected by VR systems, and (v) comparative analysis of VR and traditional methods of teaching in terms of quality of learning. This study seeks to inspire and inform interdisciplinary researchers and learners about the ways in which VR might support them and also VR software developers to push the limits of their craft.

## 1. INTRODUCTION

Virtual Reality (VR) refers to interactive scenes in which people's' viewpoint could be moved or manipulated around in a virtual three-dimension and then presented as a two-dimensional image on a computer screen. From the standpoints of communication researchers, VR is simply defined as real or simulated environment in which a perceiver experiences telepresence or an apparent participation in distant events [1]. VR is one of the many components of Information and Communication Technology (ICT) that support interactive and collaborative activities, it can transform the ways in which we learn and engage with one another. It allows for action, movement, and sometimes speech on the part of users. In theory, it may be any or a combination of visual, auditory, haptic, olfactory, gustatory, or thermal senses. The keywords to VR are visualization, which is concerned with presenting data in ways that make them perceptible and natural interaction, which allows an easier responds to the computer [2].

VR is transforming education and learning by reshaping the taught process into an interactive experience embodied in the objects of the virtual environment [3]. According to Chen et al. [4], VR offers various capabilities that can provide promising support for education. Some of these capabilities include the ability to allow the learners to experience, manipulate, and articulate their understanding of the virtual environment in real-time, interact with 3D virtual representation, and visualize abstract concepts and the dynamic relationships between several variables in the virtual environment. The system also allows individuals to collaborate with each other in a virtual environment as well as to visit and interact with events that are unavailable or unfeasible due to barriers such as distance, time, cost, or safety factors. With VR headset students can explore 3D spaces and experience dangerous, expensive or inaccessible places and events. VR can be utilized to simulate emergencies, witness volcanic activity from close-by, to walk through ancient cities, and fly through the solar system. Those studying Architecture can evaluate their buildings in new ways, pilots training can be simulated, power engineering students can simulate surge and how to control it in electric power systems, and medical students can learn about the body in 3D as shown Fig. 1 [5].

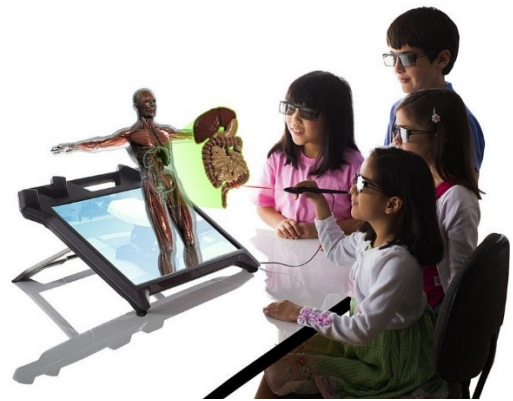

Fig. 1. The use of VR to manipulate objects in a virtual world [6]

With the continued escalation of affordable computing equipment, VR has the potential to become a powerful tool in engineering education [7]. The growing need of learner-centered education in recent years has motivated the application of virtual reality learning environment as an approach via engaging problem-solving activities [8]. This study seeks to explore the potential of VR in interdisciplinary research and education by investigating the educational scenarios and impacts of VR. The rest of the paper is structured as follows. Section 2 presents a background on the use of VR in education containing a concise review of the related literature and some popular taxonomies of VR found in the literature. Section 3 discusses the education applications of VR systems. Section 4 highlights the impacts of the use of VR over traditional approaches. Section 5 introduces the application of VR in interdisciplinary research and finally, Section 6 provides an overview of this study and a pointer to lessons learned and future work.

## 2. BACKGROUND

This section presents a concise literature review of the educational impact offered by VR systems and the taxonomies of VR found in the literature.

### 2.1 Literature Review

VR generates simulated environments to enable interactive, intuitive, and self-exploratory methods of learning [9]. Real-time interactivity is a critical feature of VR which allows learners to more effectively engage with the learning system [10]. The learner can directly interact with virtual objects, test their ideas, and observe the results in real time. The virtual environment provides an ideal problem space for learners to find their own solutions by changing the state of the objects in the virtual world. Multiple studies [11,12,13,14] have identified immersion as the key-added value of VR. Immersion creates a sense of immediacy and control through visual, auditory, or haptic devices that reflect scene changes in response to user activity. Another distinguishing feature of VR is the ability to interact with spatial representations from various frames of references (FOR), which can deepen the learner's understanding by providing different and complementary insights of the knowledge [15]. Users can develop a strong conceptual understanding of the knowledge and the relationship between its components by engaging with information from multiple roles and perspectives. VR increases motivation and mindful engagement by offering challenging, interactive, realistic, and immersive environments to learn from [13, 16]. High schools have effectively deployed VR to stimulate interest in algebra, geometry, science, and the humanities using only the crudest equipment [17]. These environments may also be artificial environments that simulate aspects of the real world which may not be accessible through direct experience [18]. For example, Immersive VR systems can model abstract phenomena (e.g., quantum mechanics) that do not have real-life referents because they cannot be experienced in practice or perceived by human senses.

MaxwellWorld (MW) is a 3-D immersive VR system [14] designed to help learners understand electrostatic concepts. These concepts are difficult to comprehend since they are abstract, three-dimensional, and lack real life referents to which learners can anchor their understanding. In MW, users can create electrostatic forces and explore electrostatic fields without other phenomena interfering with their perceived effects. Users can release positive and negative charges of varying magnitudes into the virtual environment and then interactively examine the resulting configuration. Alternatively, users can switch FOR to become a tiny charged particle enhancing the saliency of force and energy as crucial variables. Students demonstrated more in-depth understanding of electrostatic concepts and attributed immersive 3-D representations interactivity. The ability to alternate between multiple perspectives (FORs) proved to be influential to the learning process. Students reported that they thought MW was a more effective and motivating method of learning electrostatic concepts than either textbooks or lectures. Researchers evaluated the learning outcomes between MW and a highly regarded and widely used computer application called EM Field (EMF), which uses 2-D representations and quantitative values to indicate strength [19]. Pre and post lesson assessments indicated MW students developed significantly better understanding of concepts than EMF students did.

The studies by Lange and Bell et al. [5, 7, 20] identified key steps to be followed and questions to be answered in the development and implementation of VR based educational modules. Beyond just education, VR is beginning to gain traction in interdisciplinary research. This is the main contribution of this study, we want to explore the potential of VR in supporting interdisciplinary communication, which is vital for interdisciplinary education and research.

### 2.2 VR System Taxonomies

Chengling et al. [21] identified the following types of VR systems: Immersive, Desktop, Collaborative, Projected Reality, and Telepresence. In Immersive VR, the systems perceptually surround the user, replacing the sensory information from the physical world with sensory information in the virtual environment. The user is equipped with a head mounted display, trackers, and input devices that work together and are immersed through a variety of channels [22]. In a Desktop VR, the system displays a 2-Dimensional (2D) or 3-Dimensional (3D) virtual environment on a computer screen rather than a head mounted display. Despite the elimination of the immersive feature, the user can still experience the virtual world in a first-person perspective by navigating freely through the space and interacting with the virtual environment using a controller such as a joystick or a standard computer mouse. In Collaborative VR, normally implemented on the immersive or desktop VRs, the system involves a multi-user virtual environment in which participants interact with each other via avatars for instance through the tracking of each other's movements, gestures, expressions, and sounds. Projected Reality, however, provides a second-person experience in which the user stands outside, but can see themselves as part of the virtual world superimposed onto a computer monitor. A video camera serves as the input device and is used by the users to view themselves inside a computer-generated virtual world presented on a screen. Finally, Telepresence allows users at one site to control a robot or another device located at a remote site. This is achieved by using a variety of devices such as cameras, microphones, and tactile and force feedbacks to simulate the feeling of being at another location (e.g., space or undersea).

The study by Yao et al. [9] categorized current VR systems based on their features into four types: Desktop, Immersion, Reinforcement, and Network Distributed. The Desktop type utilizes PC as virtual environment generator while a computer screen is the window by which participants observe the virtual environment. The Immersion type basically utilizes high-grade workstations, high performance graphics acceleration cards and interactive devices to immerse totally into the virtual world. The Reinforcement type allows the participants to see objects in real environment, and at the same time, graphics in virtual environment are stacked on the real objects. Finally, the Network Distributed type is a large distributed interactive simulation network system composed of the previous types, which is usually used in accomplishing more complicated tasks.

Another taxonomy of VR systems was proposed by Muhanna et al. [23]; two factors that were taken into consideration in the classification include the type of technology used in building the VR system and its level of mental immersion. In terms of technology, VR systems that require special hardware facilities were categorized as Enhanced while those VR systems which do not require such facilities are said to be Basic. In terms of immersion, the authors suggested that a system that does not provide any level of mental immersion should not be considered as a VR system, they then categorized VR systems into three levels of immersion, which included (i) lowest level, which only applies to the basic VR systems (ii) higher level, experienced in partially immersive VR systems and (iii) highest level, experienced in fully immersive VR systems. Fig. 2 shows a graphical layout of the taxonomy including the examples.

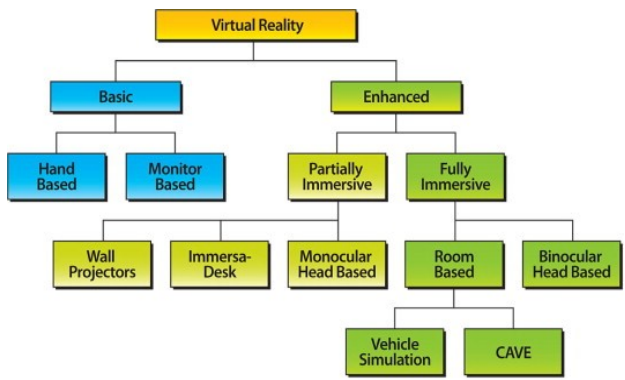

Fig. 2. Taxonomy of VR systems proposed in [23]

Yao et al. [9] categorized VR main features into (i) *multi-sensation*, which include visual sensation, auditory perception, touch perception, motion perception, taste sensation, and smell sensation; (ii) *existence sensation*, also known as immersed sense, which means the reality level that users exist in the simulated environment in the leading role; (iii) *interactivity*, which refers to the operational level where participants operate the objects in the simulated environment, and to the natural level where participants get feedback from the environment, and (iv) *autonomy*, which means the level where objects in simulated environment move independently based on the operator's requirements according to their models and regulations.

## 3. EDUCATIONAL APPLICATIONS OF VR SYSTEMS

Before substantiating the application of VR in enhancing multidisciplinary education, Chen et al. [4] first outlined some theoretical foundations which are imperative in establishing a connection with theories of learning. Forming such theoretical framework can be used to categorize the experiences and potential uses of different VRs in education [5]. The characteristics of VR, especially the immersive type, align with the axioms of constructivist learning theory, situated learning theory, embodied cognition theory, and social cognition theory.

### 3.1 Constructivist Learning Theory

According to the VR literature reviews by Chen [4] and Fokides et al. [24], constructivist philosophy holds that knowledge is constructed through an individual's interaction with the environment and that people learn better when they are actively involved in constructing knowledge in a learning-by-doing situation. VR provides a controlled environment in which learners can navigate, manipulate, and observe the effects of virtual objects found within. As such it is very well suited for providing exploratory learning environments for learning through experimentation. Both Chen and Fokides described constructivism as the underlying learning theory that fits well with the characteristics of VR technology. As such the characteristics of VR and the axioms of constructivist learning theory are entirely compatible. Chen [4] proposed a VR instructional design and development model that offers explicit guidance on how to produce an educational virtual environment based on the constructivist theory. Fokides [24] discussed the advantages and disadvantages of the use of VR in Road Safety Education, which depends heavily on the development of motor skills, and perception and time judgements based on the constructivist theoretical and cognitive perspective. One of the fundamental attributes of the multimedia and hypermedia applications that renders them compatible with the theory of constructivism is that they permit students to freely select their courses by allowing non-linear access to the contents.

### 3.2 Situated Learning Theory

Many teaching practices implicitly assume that conceptual knowledge can be abstracted from the situations in which it is learned and used, however Brown et al. [25] argued that this assumption inevitably limits the effectiveness of such practices. According to the situated learning theory proposed by Lave et al. [26], learning is situated in a specific context and embedded within a particular social and physical environment; rather than just an abstract knowledge in a classroom, learning should be embedded within activity, context and culture in which it occurs. Situated learning enables learner to takes on active role in the learning context in system being studied [8]. VR can enhance the scaffolding of situated learning through the simulation of the real world and integrating authentic tasks, thereby allowing users to learn in the context where the knowledge would be applied. Some of the features of VR that can facilitate situated learning include (i) the control and modification abilities of the environmental; (ii) the contextually rich and highly realistic nature of the learning environment; (iii) the flexibility of allowing users to adjust the difficulty of the problem, and thereby, building their knowledge and skills, and finally (iv) the opportunity for multiple practices, including interdisciplinary collaborations and practices where different variables are enforced. Brown et al. [25] described some examples of mathematics instruction (e.g. the magic square problem and story problems for teaching multiplication) that exhibit certain key features of situated learning theory to teaching. Yasin et al. [8] described an avatar implementation in VR environment using situated learning. The project was designed to guide learners through virtual reality environment as they participated in a role-based problem solving that was targeted at Hajj candidates or trainees. For those interested in many uses cases, Dawley et al. [27] provide a detailed review of recent examples of virtual worlds and immersive simulations either designed or adapted to support situated learning experiences for educational purposes.

### 3.3 Embodied Cognition Theory

Embodied cognition is the idea that the mind alone does not dictate a worldview, instead, cognition is shaped by the relationship between the mind and the body to inform and navigate the world, make meaning from the environments, and ultimately to result in learning [28]. Goldin-Meadow [29] found that simple gesturing demonstrates knowledge not found in speech and can potentially change and improve the knowledge of elementary students. Teachers use embodied cognition in the classroom whenever they instruct students to interact with the environment, for instance by conducting science experiments. Cowart [30] outlined the historical perspective of the theory. VR is one of the promising tools for delivering learning through embodied cognition. Portnoy [28] gave an illustration of an ideal high school science classroom where after reading about cells, students entered a virtual environment where they were able to push past the membrane gates and get their first glimpse of a cell and its components. Multi-sensory cues can complement embodied cognition by utilizing visual, auditory, and haptic cues in gathering information. Embodied cognition states that the knowledge we form is attributed to a broader system involving the brain, body, and environment. Users can interpret multi-sensory cues to gather information while using their proprioceptive system to navigate and control objects in the synthetic environment. This can potentially deepen learning and recall because the environment involves senses and bodily activity.

### 3.4 Social Cognition Theory

Social cognitive theory is a learning theory that is based on the idea that people learn by observing others, with the environment, behavior, and cognition acting as primary factors that influence development in a reciprocal triadic relationship. The theory revolves around knowledge acquisition or learning processes that are directly correlated to the observation of models. In other words, when learners observe a model performing a behavior and the consequences of that behavior, they remember the sequence of events and can use it to guide subsequent behaviors [31]. Although the theory originated in pure Psychology, it now finds application in Education and in Business. According to the study by Didehbani et al. [32], VR is well suited for social cognition training interventions. The authors utilized VR technology to enhance social skills, social cognition, and social functioning in young adults with autism spectrum disorders. Imam et al [33] presented another VR application in mediated learning, which was aimed to identify the VR interventions used for the lower extremity rehabilitation in stroke population and to explain their underlying training mechanisms using theoretical social cognitive and motor learning frameworks. Results obtained suggested that VR applications used mediated learning environment by facilitating task-oriented learning.

## 4. VR vs TRADITIONAL LEARNING APPROACHES

Researchers in various disciplines have studied the use of VR systems for learning. This stems from the ability of VR to implement contexts and relationships not possible to achieve in a traditional learning setting [31]. Here, we highlight the benefits of VR systems in facilitating learning over traditional methods. First, let's consider the use of VR as a tool for Road Safety teaching reported by Imam et al. [33]. Road safety education depends heavily in the development of motor skills, perception and time judgements. The course material consists of multimedia/hypermedia applications and focuses on knowledge acquisition. More so, training in conditions as close to real traffic as possible seems to be probably the only effective teaching method available. The study was an attempt to evaluate whether VR has clear advantages compared to other conventional ICT based alternatives. It concluded that VR can be a great tool of use in road safety. The main benefits that can be realized through using VR for Road Safety Education include (i) training needs to be done in a manner very close to real traffic conditions; (ii) the possibility to simulate traffic situations that are very complicated to be presented in real life, or extremely dangerous for students to be exposed to; (iii) the chance to test alternative approaches in a traffic situation through the use of elevated degree of interactions with the virtual world, and to experiment and learn from their mistakes; (iv) the fact that VR applications are very similar to the modern computer games, and as such, can stimulate students' interests by giving them extra incentives for learning, and (v) the possibility to implement different teaching styles, for example, the teacher can be present in the virtual world and can allow for collaboration between students or between a student and a guided tutoring. Unlike the real life Road Traffic education, this scenario is also perfect for interdisciplinary learning. It can allow many different professionals (health, safety, psychology) to be brought together into the learning environment. However, in VR, the students may miss, to a large extent, the benefits offered by first person experiences.

Basketball teaching has been performed for a long time by teachers interpreting and demonstrating, however, for a quick and changeable group item, the application of only the traditional teaching method can have certain limitations. VR was introduced to improve this situation and its feasibility was analyzed by Yao et al. [9] to provide a scientific reference for basketball teaching reform. Some of its benefits are as follows: (i) VR eliminates sports injury in basketball teaching; (ii) it can make up for a conditional shortage of basketball teaching and (iii) limits of room and time could be broken thoroughly.

The VR system proposed by Yasin et al. [8] showcased an immersed 3D virtual world implementing avatar for hajj trainers, some of the benefits of this VR approach over traditional training methods included the following: (i) increased understanding of the tawaf ritual; (ii) better visualization of the tawaf activity compared to the situation at the training center; (iii) flexibility, for example, having the VR system as an assistant or part of the content; while conducting the Hajj course, the trainer can simultaneously instill the learning process at the same time highlighting the participation of trainees socially during the tawaf activity; (iv) significant potential for fostering situated learning, in other words, the 3D virtual learning using the avatar allowed learners to actively immerse with the situated role based environment, and (v) interdisciplinary learning beyond tawaf activity, in other words, with the combination of high-end user graphical interface, learners could be motivated and empowered to collaborate, interact with a wide range of target learners.

For individuals with autism spectrum disorders, transition into adulthood is often a challenging time. Didehbani et al. [32] observed young adults diagnosed with high-functioning autism completed 10 sessions of social cognition training with VR technology across 5 weeks and reported a significant increase on social cognitive measures of theory of mind and emotion recognition, as well as better social and occupational functioning in real life after the training. According to the authors, previous interventions outside the realm of VR quantified social skills and cognition over time using a variety of techniques and measures and had mixed success. As such these findings suggest that the VR is a promising tool for improving social skills, cognition, and functioning in autism. Again, stroke is a global, debilitating problem which is increasing both in prevalence and incidence. Imam et al. [33] attempted to explain the underlying training mechanisms of VR interventions in stroke population based on the social cognition theory. According to the reports, results obtained demonstrated a significant improvement in outcomes in favour of the VR group compared to the control group.

Durbin [34] presented a study that was conducted by two organizations in China to take a look at the different ways VR can make public education more effective. Compared to the traditional learning approaches, VR-based learning can boast the ability to make abstract problems concrete. It is vivid, supports theoretical thinking very well, and helps sharpen students' operational skills. It makes learning more fun, secure and active by providing an immersive learning experience and enhancing students' involvement in the class. In teaching abstract courses or subjects such as astrophysics, in which learners may not be able to conduct experiments in the classroom but often will have to understand through their imagination and teachers' explanation, simulations with VR makes it possible to present to the learners the abstract aerospace in a 3D projection.

## 5. VR IN INTERDISCIPLINARY RESEARCH

*"Interdisciplinary research, education, and communication are inherently related to each other, potentially supported by complexity studies and ICT"[1]*. Interdisciplinary communication is vital to the success of every interdisciplinary research and education and VR technology is one of those ICT tools that has the potential of revolutionizing it. Multiple viewpoints are critical in interdisciplinary matters. Courses and modules are now beginning to evolve in the academia that cuts across various disciplines. For example, the principles that guided the construction of the course "Introducing Virtual Reality for Interdisciplinary Applications" in the study by Molvig [35] is the desire to mix future engineers with future non-engineers to solve problems together. Various faculties from across Vanderbilt University are serving as mentors in disciplines ranging from Art, History, and Computer Science, thereby creating interdisciplinary teams to tackle interdisciplinary projects. Other similar initiatives include the summer program for interdisciplinary research and education at the Iowa State University Virtual Reality Applications Center[2] and the VR Learning Lab[3] in Leiden, Netherlands. Hain et al. [36] discussed an industrial heritage project consisting of buildings and infrastructures where architects often found themselves in the role of mediators between investors, scientific, community, preservation, and the general public. The study found that the lack of interdisciplinary research in the project documentation resulted in non-complex solutions and low-quality developments and provided a low-level educational sense for visitors. The authors then developed a project with the aim of exploring opportunities for collaboration between theoretical research, monument preservation, virtual reality and architectural practice.

VR was utilized to represent objects and scenes within the industrial heritage conservation, and to serve as an effective communication and educational tool for rediscovering history and exploring of new scientific values by a wider range of users.

Engineers in complex automated factories often encounter communication and comprehension problems. Ulewicz et al. [37] provided a typical case study where mechanical, electrical, and computer engineers worked on a high-tech project in a team, with each party lacking an understanding of the tasks performed and complex information created by the other party in the other discipline. While their specific expertise was necessary for the project success, the integration of the different disciplines was imperative for reaching a working end result. This is particularly, a difficult interdisciplinary communication situation that VR is well suited for. The authors attempted to use VR technology to instill a mutual understanding of engineering artefacts and the connections among development teams stemming from heterogeneous domains and working at different locations. According to the report, the concept seemed very promising, yet many aspects needed to be investigated and developed in more detail.

In this section we presented some recent case studies of the use of VR in interdisciplinary research. More such systems are expected in the near future as the related technology improves to leverage VR in interdisciplinary research and education.

## 6. CONCLUSIONS

The convergence of learning theories with VR technology permits learning to be enhanced by the ability to directly manipulate objects in a virtual world. The unique affordances of VR are aligned with the components of the theories of learning on which educational applications can be developed on VR systems. Beyond just education, VR is beginning to gain traction in interdisciplinary research. This study investigates the extents to which VR technology is offering various capabilities that are able to provide promising support for education. We have also highlighted the benefits of VR systems in facilitating learning over traditional methods. Despite the potential of VR, there are still some challenges that should be addressed before implementation and deployment. We hope to look into these challenges in our future work and also develop a VR-based environment for our projects which are mostly interdisciplinary in nature. Once again, this study is intended to inspire and inform interdisciplinary researchers and learners about the ways in which VR might support them, and also VR software developers to push the limits of their craft.

## REFERENCES

1. Steuer, J., Defining virtual reality: Dimensions determining telepresence, Journal of Communication, 1992, 42(4): p. 73-93.
2. Coomans, M.K.D, Timmerman, H.J.P Towards a Taxonomy of Virtual Reality User Interfaces, Proceedings of the International Conference on Information Visualisation, 1997.

---

[1] IDREC 2018 call

[2] http://www.vrac.iastate.edu/hci/reu/

[3] http://vrlearninglab.nl/